# The Shape of the Renormalized Trajectory in the Two-dimensional O(N) Non-linear Sigma Model

WOLFGANG BOCK$^\diamond$ AND JULIUS KUTI*

Department of Physics 0319
University of California at San Diego
9500 Gilman Drive, La Jolla, CA 92093-0319

ABSTRACT. The renormalized trajectory in the multi-dimensional coupling parameter space of the two-dimensional O(3) non-linear sigma model is determined numerically under $\delta$-function block spin transformations using two different Monte Carlo renormalization group techniques. The renormalized trajectory is compared with the straight line of the fixed point trajectory (fixed point action) which leaves the asymptotically free ultraviolet fixed point of the critical surface in the orthogonal direction. Our results show that the renormalized trajectory breaks away from the fixed point trajectory in a range of the correlation length around $\xi \approx$ 3-7, flowing into the high temperature fixed point at $\xi = 0$. The analytic large $N$ calculation of the renormalized trajectory is also presented in the coupling parameter space of the most general bilinear Hamiltonians. The renormalized trajectory in the large $N$ approximation exhibits a similar shape as in the $N = 3$ case, with the sharp break occurring at a smaller correlation length of $\xi \approx$ 2-3.

# 1. Introduction

One of the most important issues in lattice field theory is the removal of cutoff effects in physical quantities. Two different approaches have recently received considerable attention, with the promise of efficient reductions in cutoff contaminations. The first approach, which combines the original Symanzik improvement program [1, 2] with tadpole improvement in the lattice coupling constant [3], has been successfully tested in the spectroscopy of heavy quark-antiquark bound states [4]. The second approach [5], which attempts to approximate the renormalized trajectory (RT) within the context of Wilson's renormalization group program [6], shows considerable promises for lattice QCD applications [7, 8]. In this work, we determine the RT in the two-dimensional O(3) non-linear sigma model and compare it with the fixed point approximation, as proposed in the second approach [5]. The precise connection between Symanzik's improvement program and a truncated approximation to the asymptotically free fixed point in Wilson's renormalization group approach remains an interesting and yet unresolved issue.

It has been known for a long time that, in contrast with the standard lattice action, each point on the RT defines a perfect lattice action [5] which is free of cutoff effects at any finite correlation length [6]. A sketch of the RT in the phase diagram of asymptotically free field theories is shown in Fig. 1 as plotted in the infinite dimensional space of couplings $K_i$. The inverse coupling $1/K_1$ of the standard lattice action is singled out to label the horizontal axis. The critical manifold is given in Fig. 1 by the $K_1 = \infty$ plane and the asymptotically free fixed point on the critical surface is designated by UVFP. The continuum theory at finite lattice correlation lengths is defined by the RT which flows along the unstable direction from the UVFP to the high temperature fixed point (HTFP) with vanishing correlation length at $K_i = 0$ [6]. The determination of the perfect lattice action has not been considered feasible in the past, mainly because a practical approximation to the RT in the infinite dimensional coupling parameter space was thought to require the uncontrolled truncation of a large number of terms in the blocked Hamiltonian.

Considerable progress has been made recently by Hasenfratz and Niedermayer [5] who realized that the fixed point lattice action of the UVFP can be determined from a classical saddle point problem in asymptotically free field theories. The fixed point trajectory (FPT), which is defined as the straight line that originates from the UVFP and is perpendicular to the critical surface, has been suggested to be a good approximation to the RT at sufficiently large correlation lengths [5] (dashed line for $K_1 < \infty$ in Fig. 1.) It was also demonstrated that the fixed point action (FPA) can be rendered very short-ranged by optimizing the block-spin renormalization group transformations with the expectation that an approximate FPT which is based on a truncated FPA, with the very small long range couplings neglected, will exhibit almost no cut-off dependence at large and moderate correlation lengths.

The performance of the approximate FPT has first been tested by Hasenfratz and Niedermayer in the 2d non-linear O(3) sigma model which is known to be asymptotically free [5]. They carried out a pilot study using a truncated FPT with 24 different couplings. Several tests showed that the residual cutoff effects were not visible even down to a correlation length of *three* suggesting that truncation effects are negligible. This indirectly implies that the FPT

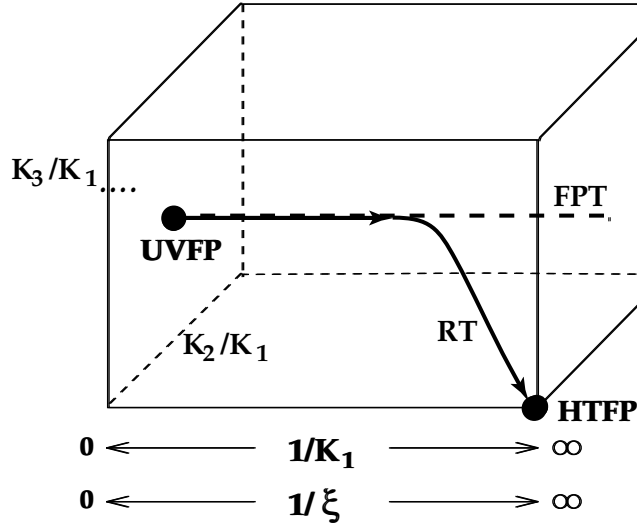

FIGURE 1. *Phase diagram of asymptotically free field theories. The inverse coupling $1/K_1$ of the standard action, or equivalently the inverse correlation length, labels the horizontal axis, and the ratios $K_i/K_1$ label the remaining (infinitely many) axes. The abbreviations RT, FPT, UVFP, and HTFP stand for renormalized trajectory, fixed point trajectory, ultraviolet fixed point, and high temperature fixed point, respectively.*

runs close to the RT in an extended range of lattice correlation lengths.

For a better understanding of the FPT approximation, it is important to determine the crossover region in the correlation length where the RT eventually has to break away from the FPT to flow into the HTFP as depicted in Fig. 1. In this work, we determine numerically the position of the RT in a finite dimensional subspace of the infinite dimensional coupling parameter space and compare it with the FPT. A more detailed account of our investigation will appear elsewhere [9].

## 2. The non-linear O(3) sigma model and the fixed point action

The path integral of the O(N) non-linear sigma model in the continuum is defined by

$$Z = \int \mathcal{D}[\phi(x)] \exp\left[-\beta \mathcal{H}_{\text{cont}}\right] , \tag{2.1}$$

$$\mathcal{H}_{\text{cont}} = \frac{1}{2} \int d^2x \, \partial_\mu \phi(x) \partial_\mu \phi(x) , \tag{2.2}$$

where $\beta = 1/T$ is the inverse temperature, $\phi(x)$ denotes the $N$-component scalar field of unit length, and $\int \mathcal{D}[\phi(x)]$ is the O(N) invariant measure. If we integrate out the momenta between the cutoff $\Lambda$ and $\Lambda/\tau$, we find the two-loop renormalization group result

$$\frac{dT}{d\ln \tau} = \frac{N-2}{2\pi} T^2 + \frac{N-2}{(2\pi)^2} T^3 + O(T^4) , \tag{2.3}$$

which implies that the model is asymptotically free for $N \geq 3$ and the UVFP is located at $T = 0$ ($\beta = \infty$.)

The lattice Hamiltonian $\mathcal{H}$ is not unique. The only constraints come from the symmetries of the system and the fact that it ought to reduce in the classical continuum limit to the Hamiltonian $\mathcal{H}_{\text{cont}}$ of Eq. (2.2). With these constraints, the lattice Hamiltonian can be parametrized as

$$\mathcal{H} = \mathcal{H}_2 + \mathcal{H}_4 + \mathcal{H}_6 + \dots , \tag{2.4}$$

$$\mathcal{H}_2 = -\frac{1}{2} \sum_r \sum_x \rho(r) \, (1 - \phi_x \phi_{x+r}) , \tag{2.5}$$

$$\mathcal{H}_4 = \sum_{x,y,z,w} c(x, y, z, w) \, (1 - \phi_x \phi_y) \, (1 - \phi_z \phi_w), \tag{2.6}$$

where $\mathcal{H}_2$ includes all terms that are bilinear in the field variable $\phi_x$, and $\mathcal{H}_4, \dots$ denote all the other terms which are quartic or higher order in the field variables. The summations in Eqs. (2.5) and (2.6) are over two-dimensional lattice vectors $r$, $x$, $y$, $z$, and $w$. The standard lattice action is given by the first two terms in Eq. (2.5) with $r_1 = (1,0)$, $(0,1)$, and $K_1 = \beta \rho(r_1)$. Physical quantities are measured in units of the lattice spacing $a$ which is set to one for notational convenience.

Let us now consider the following $\delta$-function block spin transformation,

$$\exp\left[-\beta' \mathcal{H}(\phi'; \rho', c', \dots)\right] = \int \mathcal{D}\phi \, P(\phi', \phi) \, \exp\left[-\beta \mathcal{H}(\phi; \rho, c, \dots)\right], \tag{2.7}$$

$$P(\phi', \phi) = \prod_{x'} \delta\left(\phi_{x'}' - \frac{\sum_{x \in x'} \phi_x}{\|\sum_{x \in x'} \phi_x\|}\right), \tag{2.8}$$

where a blocked lattice site $x'$ is assigned to a $2 \times 2$ cell of sites $x$ on the unblocked lattice. We put primes in Eqs. (2.7) and (2.8) to label the quantities on the blocked lattice. For details on how the FPA can be computed in asymptotically free field theories we refer the reader to Ref. [5]. The authors of Ref. [5] showed that the fixed point couplings $\rho^*(r)$ agree with the ones obtained earlier for the non-interacting model with Gaussian block spin transformation [10],

$$\rho^*(r) = \int_{-\pi}^{+\pi} \frac{d^2 p}{(2\pi)^2} \, \tilde{\rho}^*(p) \, e^{ipr} , \tag{2.9}$$

$$\tilde{\rho}^*(p) = \left[ \sum_{n_1, n_2 = -\infty}^{+\infty} \frac{1}{(p_1 + 2\pi n_1)^2 + (p_2 + 2\pi n_2)^2} \prod_{i=1}^{2} \frac{\sin^2(p_i/2)}{(p_i/2 + n_i \pi)^2} \right]^{-1} . \tag{2.10}$$

We have determined the couplings for several lattice vectors $r$ by evaluating Eqs. (2.9) and (2.10) numerically. The values of the fourteen largest couplings are given in Table 1. The interactions are very short-ranged, since the couplings decrease rapidly when the distance $|r|$ between two interacting spins grows. The couplings of the type $\rho(r)$, $r = (n, 0)$ fall exponentially with $n$, $|\rho((n,0))| \sim \exp[-1.45n]$ [5]. The authors of [5] also demonstrated that the couplings can be rendered even more short-ranged by using a soft block spin transformation. Quartic and higher order fixed point couplings can be also determined from their saddle point equation.

| $r$ | $\rho^*(r)$ | $r$ | $\rho^*(r)$ |
|-----|-------------|-----|-------------|
| $r_1 = (1,0)$ | $-3.42839$ | $r_8 = (2,1)$ | $+0.01060$ |
| $r_2 = (2,0)$ | $+0.74981$ | $r_9 = (5,0)$ | $-0.00909$ |
| $r_3 = (1,1)$ | $+0.32486$ | $r_{10} = (4,1)$ | $+0.00828$ |
| $r_4 = (3,0)$ | $-0.16871$ | $r_{11} = (3,2)$ | $+0.00502$ |
| $r_5 = (4,0)$ | $+0.03877$ | $r_{12} = (5,1)$ | $-0.00274$ |
| $r_6 = (3,1)$ | $-0.01976$ | $r_{13} = (6,0)$ | $+0.00217$ |
| $r_7 = (2,2)$ | $-0.01658$ | $r_{14} = (4,2)$ | $-0.00100$ |

TABLE 1. *The fourteen largest couplings $\rho^*(r)$.*

## 3. The renormalized trajectory in the large N limit

In the large $N$ limit, Hirsch and Shenker derived the following recursion relation for the blocked two-point function in momentum space [11],

$$G'(p') = \frac{\sum_l G(p'/2 + \pi l)\, s^2(p'/2 + \pi l)}{\frac{4}{L^2} \sum_{q',l} G(q'/2 + \pi l)\, s^2(q'/2 + \pi l)}\,, \tag{3.1}$$

$$s^2(q') = \prod_{\mu=1}^{2} \frac{\sin^2 q'_\mu}{\sin^2(q'_\mu/2)}, \tag{3.2}$$

where $l$ is a vector with components equal to 0 or 1, and $L^2$ designates the number of points on the lattice before blocking. The momentum space propagator

$$G(p) = \frac{T}{\tilde{\rho}(p) + m^2} \tag{3.3}$$

is the saddle point solution of the large N expansion (in the large N limit, $\beta$ remains finite when N is factored out from the bilinear Hamiltonian.) For a given $\beta$ and $\rho(p)$, the mass gap $m$ is determined from the gap equation

$$\frac{1}{L^2} \sum_p G(p) = 1\,. \tag{3.4}$$

Eq. (3.1) can now be iterated to determine the blocked propagators after repeated block-spin transformations. In the subspace of bilinear Hamiltonians, we can obtain the couplings $\beta\rho(r)$, $\beta'\rho'(r')$, etc. from the inverse propagators by Fourier transformation, i.e., for $r, r' \neq 0$,

$$\beta\rho(r) = \frac{1}{L^2} \sum_p \frac{1}{G(p)} e^{ipr}\,, \qquad \beta'\rho'(r') = \frac{4}{L^2} \sum_{p'} \frac{1}{G'(p')} e^{ip'r'}\,, \ldots \tag{3.5}$$

We have iterated Eq. (3.1) numerically, using two different expressions for the unblocked propagator. The first choice originates from the bilinear Hamiltonian that includes the four largest couplings in Table 1,

$$\tilde{\rho}(p) = \alpha_1 \sum_\mu 2(1 - \cos p_\mu) + \alpha_2 \sum_\mu 2(1 - \cos(2p_\mu))$$
$$+ \quad \alpha_3 \left[2(1 - \cos(p_1 + p_2)) + 2(1 - \cos(p_1 - p_2))\right] + \alpha_4 \sum_\mu 2(1 - \cos(3p_\mu))\,, \tag{3.6}$$

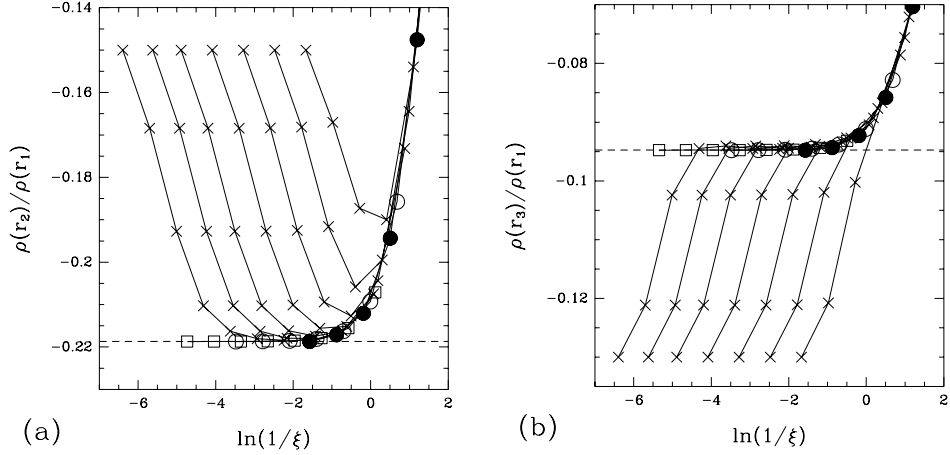

FIGURE 2. *Flow lines of $\rho(r_2)/\rho(r_1)$ (a) and $\rho(r_3)/\rho(r_1)$ (b) in the large $N$ limit. The flow lines represented by the crosses start from an initial propagator that is given by Eq. (3.6). The lines represented by squares, full, and open circles start from the fixed point propagator (cf. Eq. (2.10)). The fixed point ratios $\rho^*(r_2)/\rho^*(r_1)$ and $\rho^*(r_3)/\rho^*(r_1)$ are marked by horizontal dashed lines.*

with coupling parameters $\alpha_1$, $\alpha_2$, $\alpha_3$ and $\alpha_4$. The second choice is the fixed point propagator resulting from Eq. (2.10). In the first case, we can start the flow lines, which after several blocking steps will approach the RT, at different positions in the four-dimensional coupling parameter space by varying the five coupling parameters $\alpha_1, \ldots, \alpha_4$ and $\beta$. In the second case, the flow lines start on the FPT and it is interesting to see how the blocked couplings break away from the FPT after a few block spin transformations while tracing the exact RT. The correlation length $\xi$ on the unblocked lattice is given by $\xi = \sqrt{(\alpha_1 + 4\alpha_2 + 2\alpha_3 + 9\alpha_4)/m^2}$ for the first choice, and by $\xi = \sqrt{1/m^2}$ for the second choice (using the fact that $\tilde\rho(p)$ in Eq. (2.10) behaves as $\tilde\rho(p) \approx p^2$ for small $p^2$ values.) The parameter $m^2$ has been determined numerically from the gap equation (3.4).

As an example, we have displayed in Fig. 2 the flow lines of the ratios $\rho(r_2)/\rho(r_1)$ and $\rho(r_3)/\rho(r_1)$ as a function of $\ln(1/\xi)$. Since the correlation length is reduced by a factor of two after every block spin transformation, two consecutive points on a renormalization flow line are always separated by an amount of $\ln 2$ on the horizontal axis. The flow patterns of three-link and other couplings look very similar and we refer the reader for more details to Ref. [9]. The UVFP is located in this plot at $\ln(1/\xi) = -\infty$ and the HTFP at $\ln(1/\xi) = +\infty$. The ratios $\rho^*(r_2)/\rho^*(r_1)$ and $\rho^*(r_3)/\rho^*(r_1)$ of the fixed point couplings are represented by horizontal dashed lines. The flow lines which start from the four-coupling Hamiltonian are represented by crosses. Fig. 2 shows that those flow lines that start at large correlation lengths are attracted to the FPT within a few block spin steps indicating that the FPT is indeed a very good approximation to the RT at large correlation lengths. At small correlation lengths, however, the flow lines approach a curve (RT) which is substantially different from the FPT. The RT will eventually flow into the HTFP where the two ratios $\rho(r_2)/\rho(r_1)$ and $\rho(r_3)/\rho(r_1)$ vanish, as can be shown by the high temperature expansion [9]. Fig. 2 shows that the sharp break from the FPT occurs at a correlation length of $\xi \approx$ 2-3. The three flow lines that start on the FPT at three different values of $\beta$ are represented by squares, full, and open circles. These flow lines trace the RT quite accurately, since they start at a correlation length larger than three.

## 4. Numerical simulation of the O(3) model

The renormalization group flows in the previous section were restricted to the subspace of bilinear Hamiltonians in the large $N$ limit. In contrast, the Monte Carlo renormalization group method will allow us to study the exact flows of selected couplings in the O(3) model. Although interaction terms which do not fit on the finite lattice are truncated, their effects are expected to be negligible. The only practical limitation of the method is the signal to noise ratio in measuring very small blocked couplings.

The 1-cluster algorithm [12] has been used to generate unblocked spin configurations on a $256 \times 256$ lattice using a Hamiltonian that includes the first four couplings of Table 1. The spin configurations have been blocked four times down to a lattice of size $16 \times 16$ using the block spin transformation of Eq. (2.7). We have implemented two different methods to infer the couplings from the blocked spin configurations, a microcanonical demon method [13], and a canonical one [14]. The microcanonical method has been used earlier to determine blocked couplings in the two-dimensional O(3) sigma model [15, 16], but for a block spin transformation that differs somewhat from the one we are using in this paper. The microcanonical method performs very well from a computational viewpoint, with the disadvantage that the blocked couplings have a finite volume bias [13]. In contrast, the canonical method is exact, but suffers from larger autocorrelation effects which make the method very cumbersome from a computational point of view. For more details about the simulation and a comparison of the two techniques, we refer the reader to Ref. [9].

To illustrate our O(3) results, we have displayed in Fig. 3 the renormalization group flows of the ratios $\rho(r_2)/\rho(r_1)$ and $\rho(r_3)/\rho(r_1)$ as a function of $\ln(1/\xi)$. The correlation length on the unblocked lattice has been determined from the propagator in coordinate space by using the improved estimator technique of ref. [17]. Most of the renormalization group flow lines exhibited in Fig. 3 have been obtained with the microcanonical method (crosses.) We have also included renormalization group flow lines (circles) which were generated by the canonical demon method. For one of the canonical flow lines, the initial couplings were chosen to coincide with the starting point of a microcanonical flow line. Fig. 3 shows that the two flow lines with the same initial couplings coincide within the error bars, and the flow pattern is very similar to the one seen in Fig. 2. The only difference is that the sharp break from the FPT occurs now at a somewhat larger correlation length. The grey lines in Fig. 3 mark the approximate position of the RT, as determined from the flow lines. For comparison, we have also included the RT in the large $N$ limit (heavy lines) (cf. Fig. 2.)

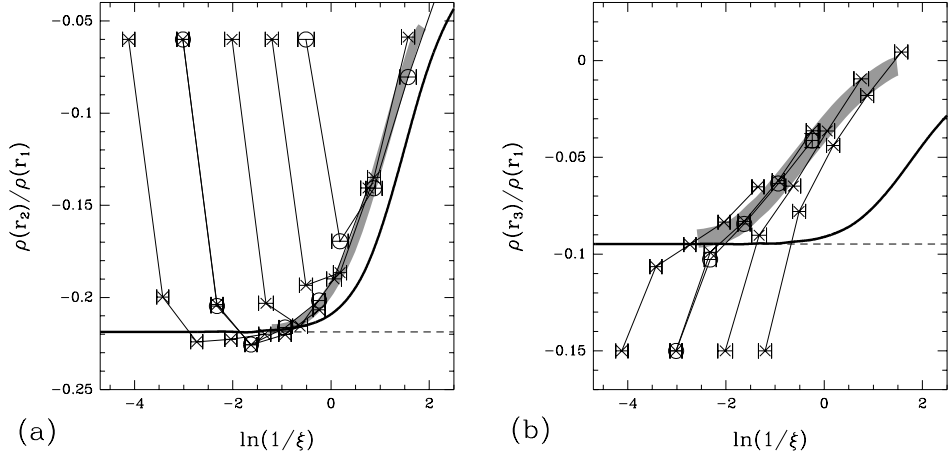

FIGURE 3. *Flow lines of $\rho(r_2)/\rho(r_1)$ and $\rho(r_3)/\rho(r_1)$ for $N = 3$. All flow lines start from the four-coupling Hamiltonian. The lines which are represented by crosses (circles) have been generated with the microcanonical (canonical) method, respectively. The grey lines indicate the approximate position of the RT, while the solid lines depict the projections of the large $N$ RT onto the two planes. The fixed point ratios $\rho^*(r_2)/\rho^*(r_1)$ and $\rho^*(r_3)/\rho^*(r_1)$ are marked by horizontal dashed lines. Error bars are omitted when they are smaller than the symbol sizes.*

## 5. Conclusion

We have demonstrated, by explicitly computing the RT in the O(3) model, that the FPT provides a good approximation to the RT in the region of the coupling parameter space where the correlation length is large. A significant break occurs, however, in a range of the correlation length around $\xi \approx 3$-$7$ where the RT sharply departs from the FPT and flows into the HTFP. The RT in the large $N$ limit exhibits a similar behavior with the sharp break shifted to somewhat smaller correlation length. Although we have only shown the projection of the exact RT in the $K_1$-$K_2$ and $K_1$-$K_3$ planes, several other projections of the flows were determined as well [9].

Both ends of the RT can be determined without the utilization of the Monte Carlo renormalization group. The RT in the large correlation length regime, near the UVFP, is well approximated by the FPT, while the RT in the regime of small correlation lengths, near the HTFP, can be obtained from a high temperature expansion. It is expected that both approximations will break down in the crossover region where the Monte Carlo renormalization group technique may remain the only useful tool. It is not clear whether this regime will be important in practical applications.

The studies we presented here will be useful to extend to non-Abelian lattice gauge theories in four dimensions [7].

*Note added.* After the completion of this work, M. Okawa pointed out some earlier results [18] where a similar large N approximation was applied to block spin transformations that differ somewhat from the ones used in our paper.

## Acknowledgements

This work was supported by the DOE under grant DE-FG03-91ER40546. One of us (W. B.) would like to acknowledge discussions with W. Bietenholz and U. Wiese.